\documentclass{iau}

\usepackage{amsmath}
\usepackage{graphicx}
\usepackage{multirow}
\usepackage[english]{babel}
\begin{document}

\lefttitle{Cambridge Author}
\righttitle{Proceedings of the International Astronomical Union: \LaTeX\ Guidelines for~authors}

\jnlPage{1}{7}
\jnlDoiYr{2021}
\doival{10.1017/xxxxx}

\aopheadtitle{Proceedings IAU Symposium No. 365}
\editors{Alexander Getling \&  Leonid Kitchatinov, eds.}

\title{Magnetic field and  radial velocity fluxes at the initial stages of the evolution of solar active regions based on measurements at the photospheric level}

\author{A. M. Sadykov and S. A. Krasotkin}
\affiliation{\textbf{Department} of Physics, Moscow State University, Moscow, Russian Federation}

\begin{abstract}
In this article, the physical processes occurring in the convective layer and the photosphere of the Sun and their connection to the formation of active regions (ARs) and the development of the corresponding magnetic field are explored. Specifically, we test the magnetic flux emergence hypothesis and based on the line-of-sight magnetic field and Doppler shift data obtained from the Global Oscillation Network Group (GONG) observations. The study encompasses the analysis of 24 ARs observed during the period from 2011 to 2022. We find a strong correlation between the magnetic flux and the imbalance of radial velocity fluxes. The results indicate that the magnetic flux emergence hypothesis cannot fully explain the evolution of ARs during their early stages of development.
\end{abstract}


\maketitle

\section{Introduction}
As of today, there are no universally accepted views on the physical processes occurring in the convective layer and the photosphere of the Sun related to the emergence of active regions (ARs) and corresponding magnetic field development. One of the most common physical theories regarding the evolution of ARs is the hypothesis of magnetic flux emergence, or alternatively, the emerging magnetic flux tube hypothesis \citep{parker1955}. This concept of emerging magnetic flux tubes is the default assumption in a significant number of theoretical works. Consequently, there is an evident need to verify the accuracy of this concept based on solar magnetic field and Doppler observations.

According to the emerging magnetic flux tube model, the ARs appear in a certain area of the surface because the magnetic flux tube loop rises from the convective layer to the photosphere due to magnetic buoyancy. According to calculations by \citet{caligari1995}, the radial velocity of the rising part of the magnetic loop increases as it moves closer to the surface. At the level of the photosphere, the radial velocity of the magnetic tube is expected to exceed $\approx$500\,m/s. Such a velocity is greater than the radial velocity of the plasma motion at the center of a granule, which is approximately $\approx$300\,m/s \citep{zirin1966solar}. Consequently, if the emerging magnetic flux tube hypothesis is correct, a local increase in the radial velocity flux of matter should be observed on the photosphere, accompanied by an enhancement of the magnetic field.

\section{Observations and data processing}

To conduct the research, we selected 24 ARs observed during the period from 2011 to 2022. The properties of their appearance on the Sun and subsequent development were studied through the analysis of a series of line-of-sight magnetograms and Dopplergrams by the network of ground-based Global Oscillation Network Group (GONG) telescopes, with a temporal resolution of 4 hours over a period of 5-6 days. Each of the selected ARs initially resided on the eastern hemisphere of the Sun and later evolved into a distinct structure consisting of two or more spots.

\begin{figure}[h]
  \centering
  \includegraphics[scale=.49, clip]{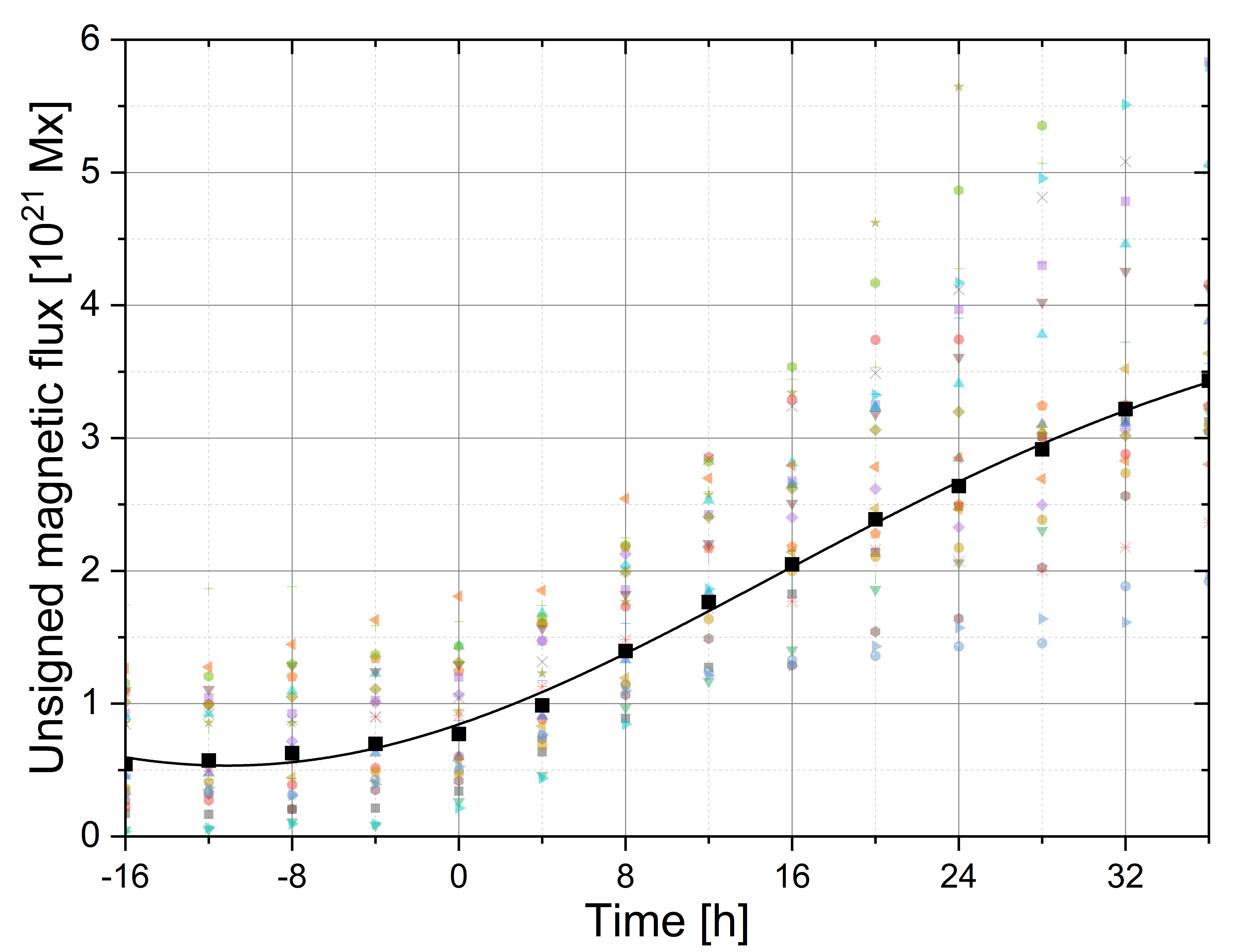}
  \caption{Unsigned magnetic flux in a unified time reference frame. The moment t=0 hours corresponds to the moment of sharp increase of the magnetic flux.}
  \label{Fig1}
\end{figure}

The time series data for all of the examined ARs were synchronized to a common time reference textbf{frame} using an epoch overlay method, ensuring that the onset of the sharp growth of the magnetic flux corresponds to t$=$0\,h of the AR evolution time. The moment t$=$0\,h was determined as the earliest time moment there the following condition is satisfied:

\begin{equation}
J_{mag}(t_0+8\:h.) - J_{mag}(t_0)\geq 0.5 * 10^{21}\:Mx\text{,}\label{eq1}
\end{equation}

where $J_{mag}$ is the unsigned magnetic field flux over the selected frame of the Solar disk (4'~*~4'), which is calculated by multiplying the magnetic field strength value in each pixel by the area of the photospheric corresponding to this pixel. Thanks to the alignment in a unified time reference system, it became possible to perform averaging for each moment of the considered magnetic field characteristics. The evolution of the unsigned magnetic flux averaged over the studied regions is presented in Figure~\ref{Fig1}.

For demonstrating the relationship between the fluxes of the upward and downward radial velocity of matter, we consider a quantity known as the imbalance of radial velocity fluxes. This quantity is defined as the difference between the fluxes of upward and downward radial velocity, normalized to their sum:

\begin{equation}
D = \frac{|J_+| - |J_-|}{|J_+| + |J_-|}\text{,}\label{eq2}
\end{equation}

where $J_+$ and $J_-$ represent the fluxes of the upward and downward radial velocity of matter integrated over the selected area, respectively. The fluxes of radial velocity are calculated in a similar manner to the magnetic field flux. As evident from the definition, the imbalance of radial velocity fluxes is dimensionless (does not require adherence to a specific system of units) and ranges from $-1$ to $1$ or from -100\% to 100\% (percentage values are used in this work).

\section{Results}

As can be seen from Figure~\ref{Fig1}, the magnetic field flux basically monotonically increases with time. Moreover, it vividly illustrates the method of superimposing epochs of AR data sampling at the moment of a sharp increase in the magnetic flux (Eq.~\ref{eq1}), which corresponds to the time label at $t_0 = 0\: \text{h}$ in a unified time reference frame.

\begin{figure}[h]
  \centering
  \includegraphics[scale=.49, clip]{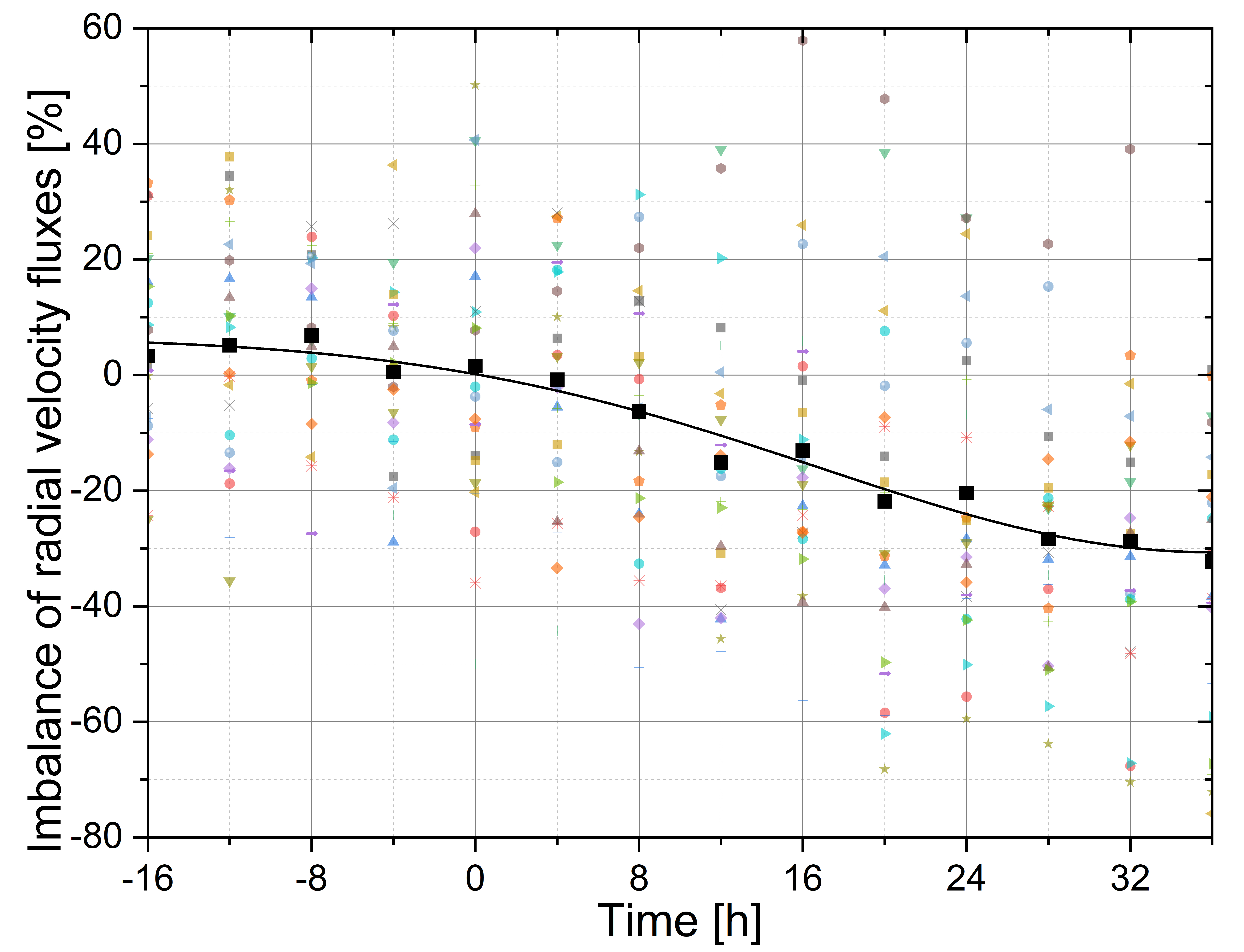}
  \caption{Imbalance of radial velocity fluxes as a function of time in a unified time reference frame.}
  \label{Fig2}
\end{figure}

The graph of the radial velocity flux imbalance over time (Fig.~\ref{Fig2}) shows a strong descending trend over time. Before the moment of the sharp increase in the magnetic field flux corresponding to $t_0 = 0\:\text{h}$ the imbalance of fluxes was around zero, indicating an equilibrium between the upflow and downflow of material in the system. Then, after $t_0 = 0\:\text{hours},$ a monotonic shift of the flux imbalance towards negative values is observed. This means that in the studied ARs, the downward motions of the material start to dominate over the upward motions.

\begin{figure}[h]
  \centering
  \includegraphics[scale=.49, clip]{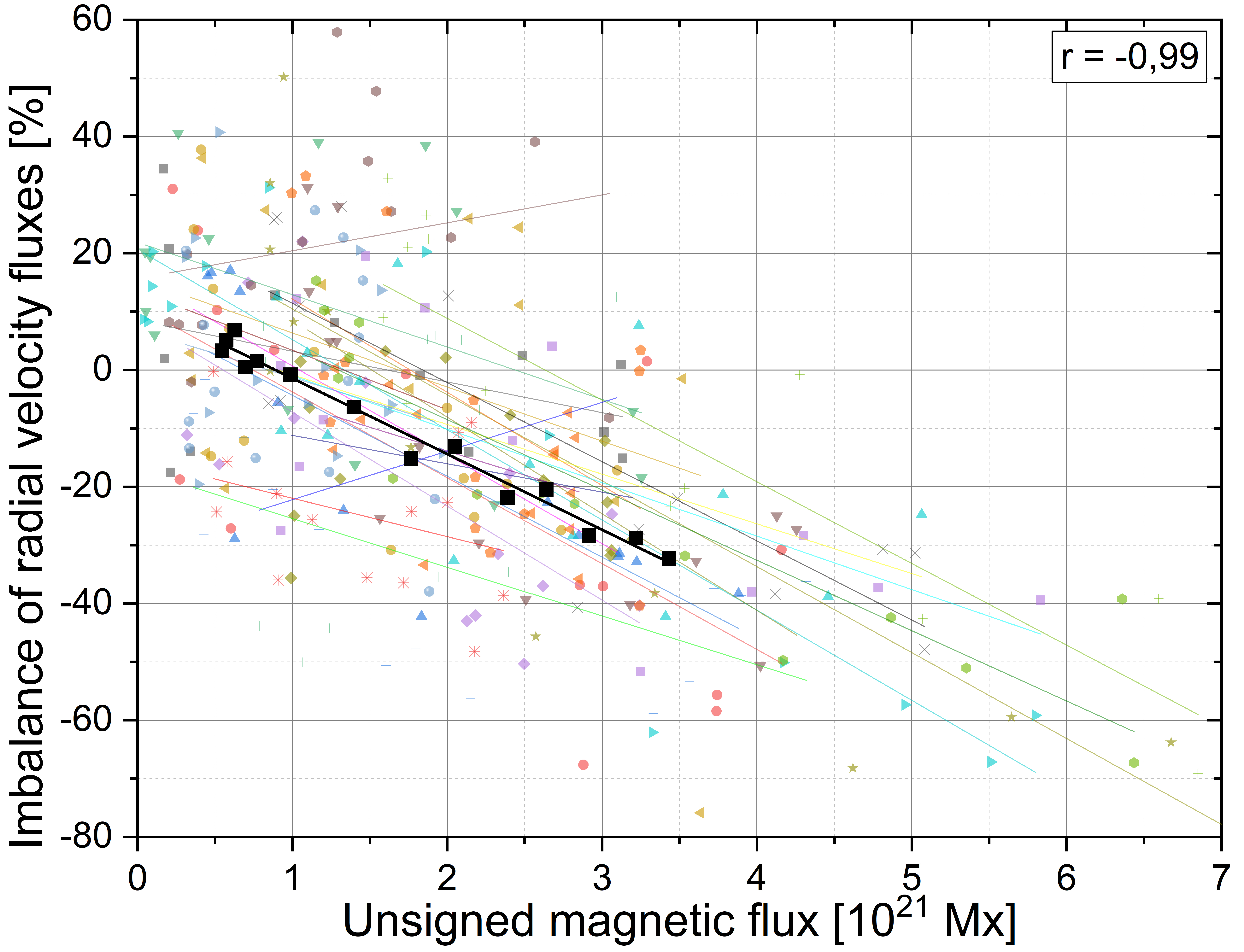}
  \caption{Imbalance of radial velocity fluxes as a function of the magnetic flux.}
  \label{Fig3}
\end{figure}

By using the binary relationship graphs, one can visually observe how the velocity flux imbalance changes with different values of the magnetic field flux. For the radial velocity flux imbalance (Fig.~\ref{Fig3}), there is a notably strong correlation with the total magnetic field flux. For averaged values, the correlation coefficient is r$=-0,99$. This implies that as the magnetic field flux increases, the material's downward velocity flux becomes increasingly dominant. At low values of the magnetic field flux, the velocity flux imbalance remains close to zero, indicating no tendency towards upward or downward material movement.

\section{Conclusion}

Based on the obtained results, it can be concluded that, on average, in the early stages of active region (AR) evolution, there is a change in the radial velocity flux of material. A sharp transition is observed from an equilibrium between the upflow and downflow radial velocity to the dominance of the downflow radial velocity as the magnetic field flux increases. The hypothesis of magnetic flux emergence implies an increase in the magnetic field flux synchronized with the increase in the upflow radial velocity. In terms of the velocity flux imbalance, this would lead to a shift towards positive values. However, in the examined ARs, no corresponding synchronous changes potentially corresponding to the emerging magnetic flux model were observed. Therefore, the magnetic flux tube emergence model should be used with caution when describing the evolution of AR, at least in the initial phases of their development, indicating a local rather than a global nature of the processes occurring during the AR formation phase.

\bibliographystyle{aasjournal}
\bibliography{iauguide}

\end{document}